\documentclass{optica-article}

\journal{opticajournal} 

\articletype{Research Article}

\usepackage{lineno}
\linenumbers 
\usepackage{xcolor}
\hypersetup{colorlinks=true,linkcolor=blue}
\usepackage{booktabs}
\usepackage{tabularx}
\usepackage{array}
\usepackage{makecell}
\usepackage{threeparttable}

\newcolumntype{Y}{>{\raggedright\arraybackslash}X}
\newcolumntype{C}{>{\centering\arraybackslash}X}
\nolinenumbers
\begin{document}

\title{Octave bandwidth 3D-Printed Couplers for Low-Loss Thin-Film Lithium Tantalate Circuits}

\author{Erik Jung\authormark{1,$\dag$}, Xinyu Ma\authormark{1,$\dag$}, Jan Brandes\authormark{1},Caghan Ünlüer\authormark{1,2}, Shabnam Taheriniya\authormark{1}, Seongmin Jo\authormark{1,2}, Philipp Schmidt\authormark{1}, Eric Biedert\authormark{1}, Liam McRae\authormark{1},  Julian Rasmus Bankwitz\authormark{1}, Frank Brückerhoff-Plückelmann\authormark{1} and Wolfram Pernice\authormark{1,2,*}}

\address{\authormark{1}Heidelberg University, Kirchhoff Institute for Physics, Im Neuenheimer Feld 227, Heidelberg, Germany.\\
\authormark{2}University of Münster, Center for Nanotechnology, Heisenbergstraße 11, Münster, Germany.\\}

\email{\authormark{*}wolfram.pernice@kip.uni-heidelberg.de} 


\begin{abstract*} 
Low-loss, broadband photonic integrated circuits (PICs) are critical enablers for optical communications, photonic computing, and quantum applications. Lithium tantalate on insulator (LTOI) is an emerging photonic platform offering a wide transparency window and strong Pockels effect, and thereby enabling efficient electro-optic modulation and high data rates.  Here, we present the first implementation of efficient out-of-plane polymer coupling interfaces fabricated via 3D direct laser writing for both fully etched strip and partially etched rib LTOI waveguides, achieving ultra-low coupling losses of 0.9 dB (strip) and 1.25 dB (rib) per interface. Both coupler types exhibit a 3 dB optical bandwidth spanning more than an octave from 850 nm to 1740 nm and maintain stable operation under 1 W optical input power. Combined with on-chip waveguides exhibiting propagation losses below 0.1 dB/cm, these characteristics represent a key step toward unlocking the full potential of LTOI for high-speed optical signal processing with an unprecedented degree of parallelism. In addition, the octave-spanning bandwidth enables efficient interfacing of both the fundamental and second-harmonic signals, making the platform highly attractive for second harmonic generation based quantum squeezing applications.

\end{abstract*}
\section{Introduction}

Low-loss and broadband photonic integrated circuits (PICs) are essential for energy-efficient photonic devices and therefore determine the scalability of optical communication systems \cite{zhang_integrated_2026,broadband_eom} and photonic computing architectures\cite{bandyopadhyay_single-chip_2024, shen_deep_2017,meyer_deep_2026,bruckerhoff-pluckelmann_uncertainty_2025}. They are also a key enabler for quantum photonic experiments\cite{zhengLargescaleQuantumCommunication2026, larsenIntegratedPhotonicSource2025}, since optical losses fundamentally limit quantum state coherence and entanglement strength \cite{spdc_jo, squeezer_zhu}. 

 Established PIC platforms each offer distinct advantages but also involve trade-offs between optical loss, modulation speed, transparency range, and material stability. Si$_3$N$_4$ enables ultra-low-loss waveguides and broad transparency but relies on slow thermo-optic modulation\cite{ravi_sin,xu2026_trim}. SOI supports GHz-speed carrier-based modulation at the cost of additional absorption and limited transparency toward visible wavelengths\cite{si_review}. LNOI offers low-loss waveguiding \cite{loncar_ln_aplp} and high-speed Pockels-based modulation \cite{cai_2019np} but exhibits strong birefringence and photorefraction effect\cite{tang_pr_ln}. Thin-film lithium-tantalate-on-insulator (LTOI) extends the transparency window into the ultraviolet\cite{lt_uv} and provides higher optical damage thresholds with reduced birefringence\cite{dai_as, lt_damage_threshold}, making it attractive for low-loss, broadband and high-power photonic applications\cite{tjk_lt_nature_2024}.

However, these advantages can only be fully unlocked using coupling interfaces with comparable high coupling efficiency, bandwidth and power tolerance. Conventional grating couplers allow for wafer-scale testing, but typically exhibit limited bandwidths of a few 10s of nanonmeter\cite{ma_oe, emma_gc}. In-plane couplers such as spot-size converters provide broader bandwidth at the costs of routing constraints\cite{table_ref_5}. In contrast, three-dimensional polymer out-of-plane couplers combine sub-dB loss and ultra-broadband coupling  with high design flexibility\cite{gehring_low-loss_2019,Jung_Broadband_25}. Previous demonstrations, however, have been limited to fully etched strip waveguides on passive Si$_3$N$_4$ platforms, whereas active electro-optic devices require partially etched rib waveguides for efficient modulation\cite{dai_as}. Due to the lack of octave-spanning coupling bandwidth, second-harmonic-generation (SHG)-based squeezing systems either require separate couplers for the fundamental and second-harmonic wavelengths\cite{lohmann} or suffer from a significant difference in coupling efficiency between the two wavelengths, which quadratically impacts the achievable squeezing levels\cite{zhao_shallow-etched_2020}.

Here we present a scalable low-loss out of plane coupling interfaces for both fully-etched strip as well as partially-etched rib LTOI waveguides in combination with low-loss waveguides. 
We demonstrate coupling losses as low as 0.9 dB per strip waveguide coupler and 1.25 dB per rib waveguide coupler, with a 3 dB bandwidth exceeding an octave for both designs.
Additionally, we achieve low-loss on-chip waveguides with propagation losses of 0.11 and 0.06 dB/cm in the single-mode and multi-mode regimes, respectively. Furthermore, we verify high-power operation tolerance by guiding 1 W of input power at a wavelength of 1550 nm through both types of LTOI circuits.
The high bandwidth, low-loss and, high power damage threshold make LTOI interfaced with 3D printed coupler a prime candidate for highly parallel electro-optic modulation and pave the way for quantum photonic networks based on squeezing. 
\begin{figure}[htbp]
\includegraphics[width=\linewidth]{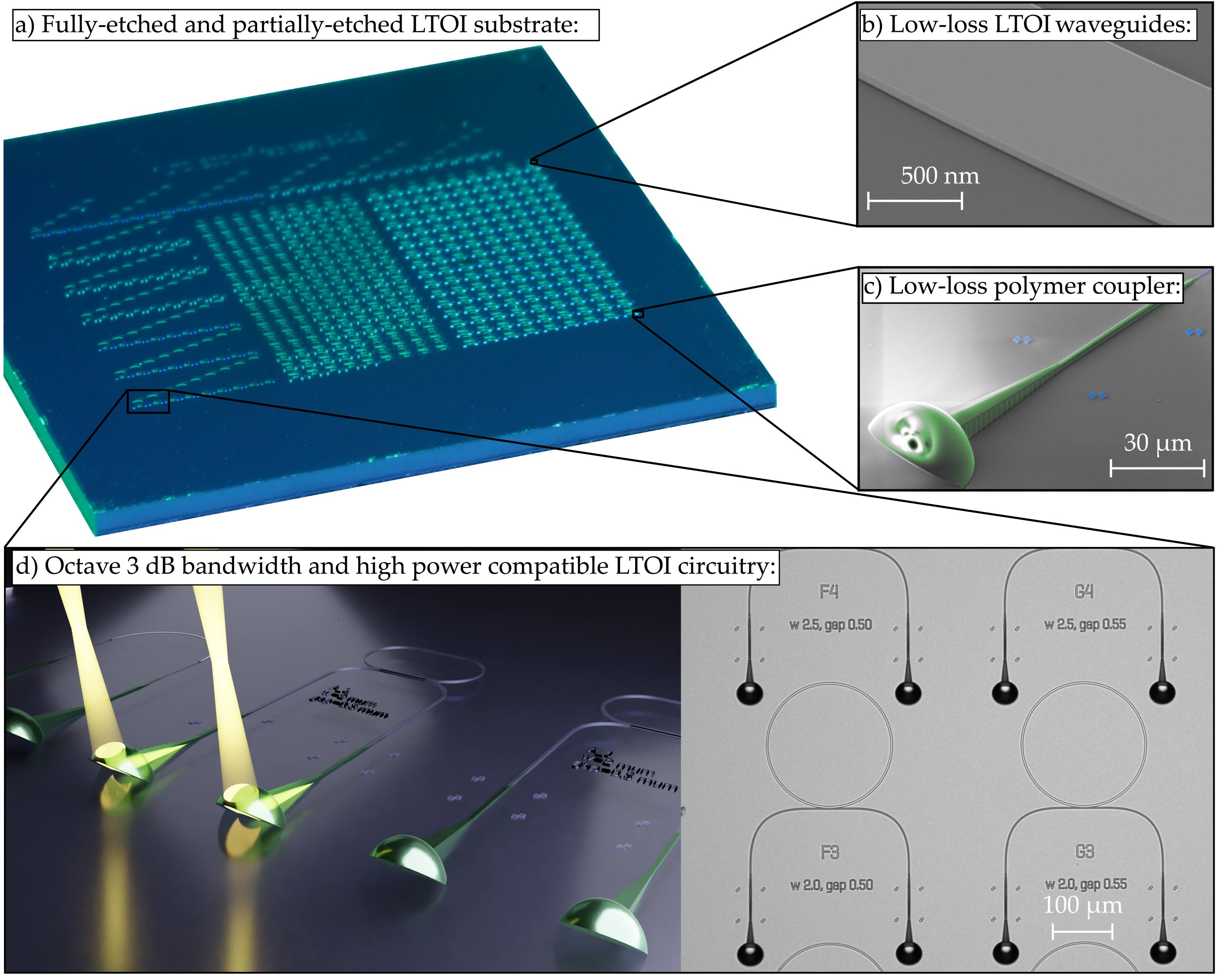}
\caption{\textit{Building blocks for LTOI circuitry:} (a) Photograph of a LTOI photonic chip, with insets highlighting individual on-chip components. (b,c) Scanning electron microscope (SEM) images of low-loss LTOI waveguides integrated with polymer TIR couplers (highlighted in green) for fully etched strip waveguides and partially etched rib waveguides. (d) Combination of these components enabling low-loss, octave-bandwidth, and high-power-compatible on-chip circuits, as shown in the conceptual image and corresponding microscope image. }
\label{fig1}
\end{figure}
\section{Results}

The fundamental building blocks of both fully etched and partially etched LTOI chips, shown in the photograph in Fig. \ref{fig1}(a), are low-loss on-chip waveguides and high-efficiency coupling interfaces. Close-up views of these components are presented in the scanning electron microscope (SEM) images in Fig. \ref{fig1}(b) and (c), highlighting the high structural quality of the waveguides and the low-loss polymer coupler, marked in green.

By combining these building blocks, we realize octave bandwidth, low-loss and high optical damage threshold passive circuits. Fig. \ref{fig1}(d) shows a conceptual image and the corresponding microscope image of a passive ring resonator interfaced with 3D-printed couplers. The couplers, highlighted in green, redirect the optical beam out of the chip plane and are aligned via on-chip markers to ensure precise positioning relative to the waveguides.
\begin{figure}[htbp]
\includegraphics[width=\linewidth]{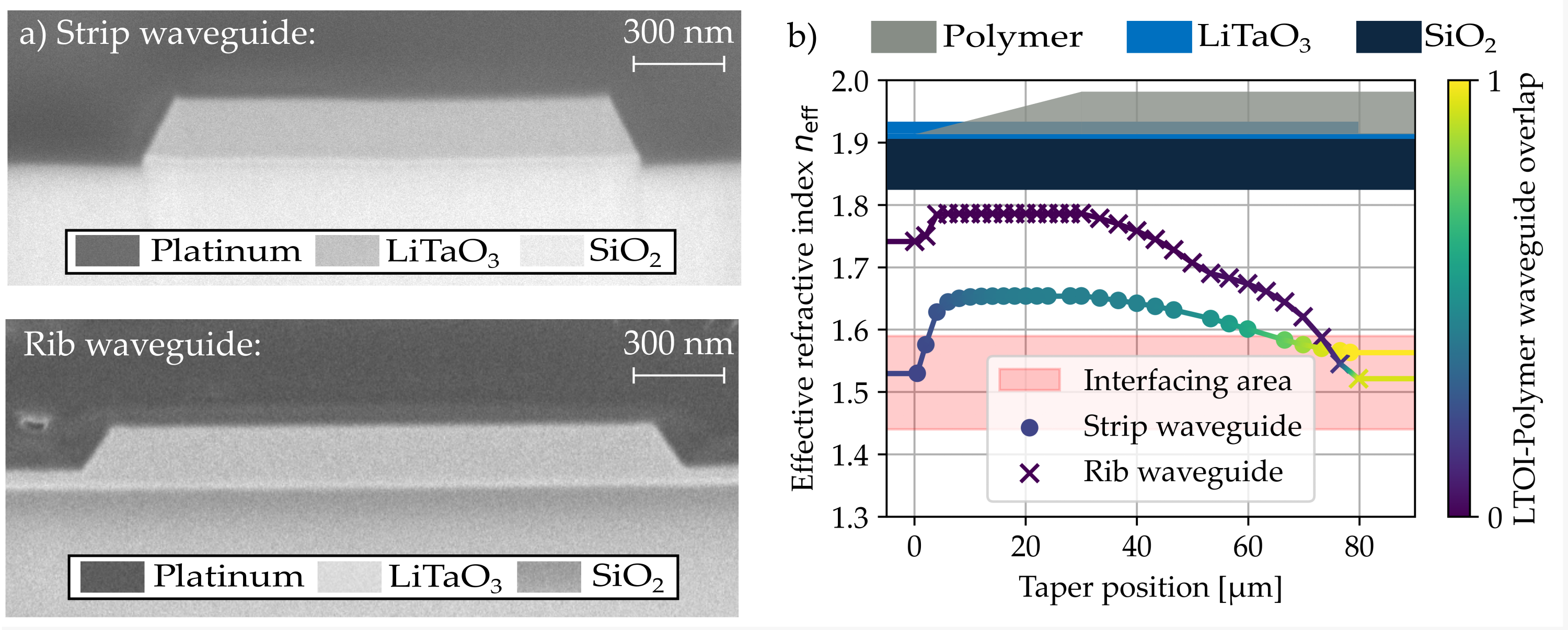}
\caption{ \textit{LTOI-polymer interface:}(a) Scanning electron microscope (SEM) images of focused ion beam (FIB) cutted cross-sections of fully etched strip waveguides and partially etched rib waveguides. (b) Evolution of the effective refractive index along the mode-transfer taper at 1550 nm for both waveguide types, illustrating the efficient transition from LTOI-like mode to a polymer waveguide mode, as quantified by the color-coded overlap integral.}
\label{fig2}
\end{figure}

On-chip waveguides can be categorized into fully etched strip and partially etched rib geometries. Fig. \ref{fig2}(a) presents focused ion beam (FIB) cut cross-sections of both types. Strip waveguides are etched down to the $\mathrm{SiO}_2$ substrate, providing strong optical confinement and enabling a reduced device footprint. In contrast, rib waveguides are only partially etched yielding in a residual $\mathrm{LiTaO}_3$-waveguide slab, a geometry that is particularly advantageous for electro-optic modulation. The presented waveguides are fabricated on a 200 nm-thick LTOI platform with a etching depth of $140\,\mathrm{nm}$ for the rib waveguides. For imaging, the cut cross-sections are coated with platinum, which is else not present. The characteristic side-wall angle for the fully etched strip waveguide for the partially-etched rib waveguide are observed.

To fully exploit the large transparency window of LTOI e.g. in the context of large-scale wavelength-division multiplexing, the waveguide circuits must be interfaced with broadband couplers. Three-dimensional printed total internal reflection (TIR) couplers are prime candidates for this purpose, offering an ultrawide transparency window.

In order to interface the high refractive index on-chip waveguides ($n_{\mathrm{LiTaO}_3} (\lambda=1550\,\mathrm{nm})= 2.12$) with the lower refractive index couplers ($n_{\mathrm{poly}} (\lambda =1550\,\mathrm{nm})= 1.52-1.59$), the waveguide is tapered down while a single-mode polymer waveguide is printed on top. As the taper narrows, the effective refractive index of the guided mode decreases and falls below the refractive index of the polymer waveguide material, causing the modal overlap with the polymer mode to increase. This enables the mode to couple adiabatically from the on-chip waveguide to the 3D printed polymer waveguide. A schematic of this transition region, along with the evolution of the effective refractive index at 1550 nm determined by 2D finite element frequency domain (FEFD) simulations of a cut through the cross-section, is shown in Fig. \ref{fig2}(b).

For the fully etched strip waveguide, the resin IPX-clear (Nanoscribe GmbH) is selected due to its high transparency window and low absorption down to the visible regime. The color coded overlap integral of the  $\mathrm{TE}_0$ mode of the LTOI and the polymer waveguide is strongly increasing after the on-chip waveguide has reached a width smaller than 300 nm, indicating the transition of the mode from the on-chip waveguide to the polymer.  
For partially etched rib waveguides, simulations indicate that the optical power would couple into the remaining slab rather than to the polymer waveguide due to slab possessing a higher refractive index than the IPX-clear printed polymer waveguide atop. In order to prevent the coupling of the mode to the slab, for the rib waveguide coupler the resin IPN-162 (Nanoscribe GmbH) with a refractive index 

\begin{equation}
    n_\mathrm{IPN-162}(\lambda=1550nm)\approx1.59 > n_\mathrm{IPX-clear}(\lambda=1550nm)\approx1.525
\end{equation}
is used. The colorcoded overlap integral starts to increase at the very tip of the on-chip waveguide taper, showing that the slab waveguide supports a propagating mode even with cross-sections in the sub 100 nm regime. This overlap distribution marks the refractive index of IPN-162 as the minimum refractive index to interface rib waveguides with polymer structures. A simulation of the evolution of the mode field in the mode transition section is depicted in the supplementrary material.

\begin{figure}[h!]
\includegraphics[width=\linewidth]{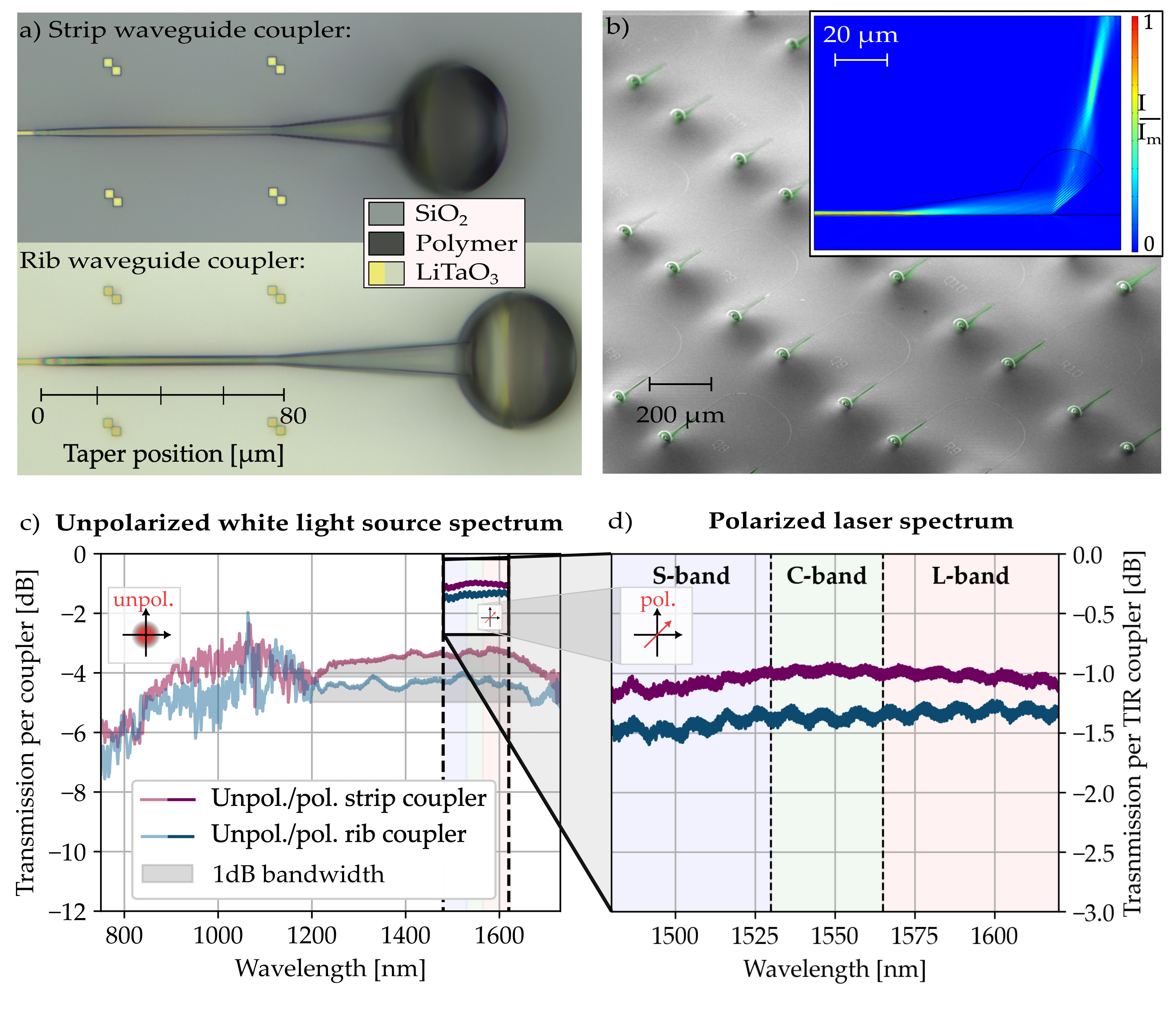}
\caption{\textit{Coupler performance: }(a) Microscope images at identical scale showing the different polymer coupler designs for strip  and rib waveguides. (b) SEM image of multiple couplers on an LTOI chip with inset showing the simulated intensity evolution in a rib coupler. (c) Transmission spectrum from 750 nm to 1730 nm obtained with a white light source showing the octave spanning 3 dB bandwidth and the 1 dB bandwidth exceeding 450 nm for both coupler types. The inset shows the transmission spectrum from 1480 nm to 1620 nm of the couplers  measured with a polarized tunable laser. (d) Close-up of the nearly wavelength independent spectrum for both coupler types. The deviation between the unpolarized and polarized spectra is caused by the difference in polarization.}
\label{fig3}
\end{figure}
Due to the different materials of the couplers, the TIR couplers for the strip and rib waveguides differ in form. Microscope images of the two different couplers are depicted in Fig. \ref{fig3}(a). For fully etched substrates, the underlying $\mathrm{SiO}_2$ layer is visible, whereas for partially etched waveguides, the residual slab layer covers the entire chip, leading to a different visual appearance of the chips.
Each coupler consists of four main sections. The coupler starts with a mode transfer taper, where the mode is transferred from the on-chip waveguide to the polymer waveguide. It is followed by a mode-field expansion taper, adapting the mode size to match that of the fiber after free-space propagation. A total internal reflection surface is redirecting the beam path out of the chip's plane. To match the numerical aperture of the fiber's beam, a focusing lens is used. Due to the high-refractive index slab, the mode of the ridge waveguide coupler is expanding slower in comparison with the rib waveguide coupler. Consequently, these couplers require a longer mode-field expansion taper and a less steep TIR angle.

An SEM image of multiple TIR couplers fabricated on the LTOI platform is shown in Fig. \ref{fig3}(b) demonstrating compatibility with large-scale fabrication. The inset illustrates the simulated evolution of the intensity profile $I$ normalized to the maximum intensity $I_\mathrm{m}$ within a coupler on a partially etched substrate. A comparison of the evolution of the intensity profiles within the two coupler types is depicted in the supplementary information. 

Using an unpolarized white light source in combination with an optical spectrum analyzer, the transmission spectra of the couplers over a wavelength range from 750 nm to 1730 nm are measured. The results are depicted in Fig. \ref{fig3}(c). The white light sources output of 150 mW distributed over the whole spectrum from 450 nm to 2400 nm is attenuated with a 20 dB attenuator. Both fully etched strip and partially etched rib designs exhibit ultra-broadband transmission, with 1 dB bandwidth thresholds of 485 nm and 469 nm respectively and a 3 dB bandwidth exceeding one octave.  The pump light of the supercontinuum white light source is at 1064 nm, causing the spectrum to be more noisy around the pump.
Additionally, the transmission spectra are measured in the telecom wavelength range using a polarized tunable laser from 1480 nm to 1620 nm. These measurements are shown as an inset in Fig. \ref{fig3}(c), while Fig. \ref{fig3}(d) presents a close-up view of the same data.
 Spanning over the telecom S-band , C-band and L-band, both the couplers show a low-loss and nearly wavelength independent transmission spectrum. In particular, the strip waveguide coupler exhibit a peak transmission of -0.9 dB and the rib waveguide coupler of -1.25 dB.

\begin{figure}[h!]

\includegraphics[width=\linewidth]{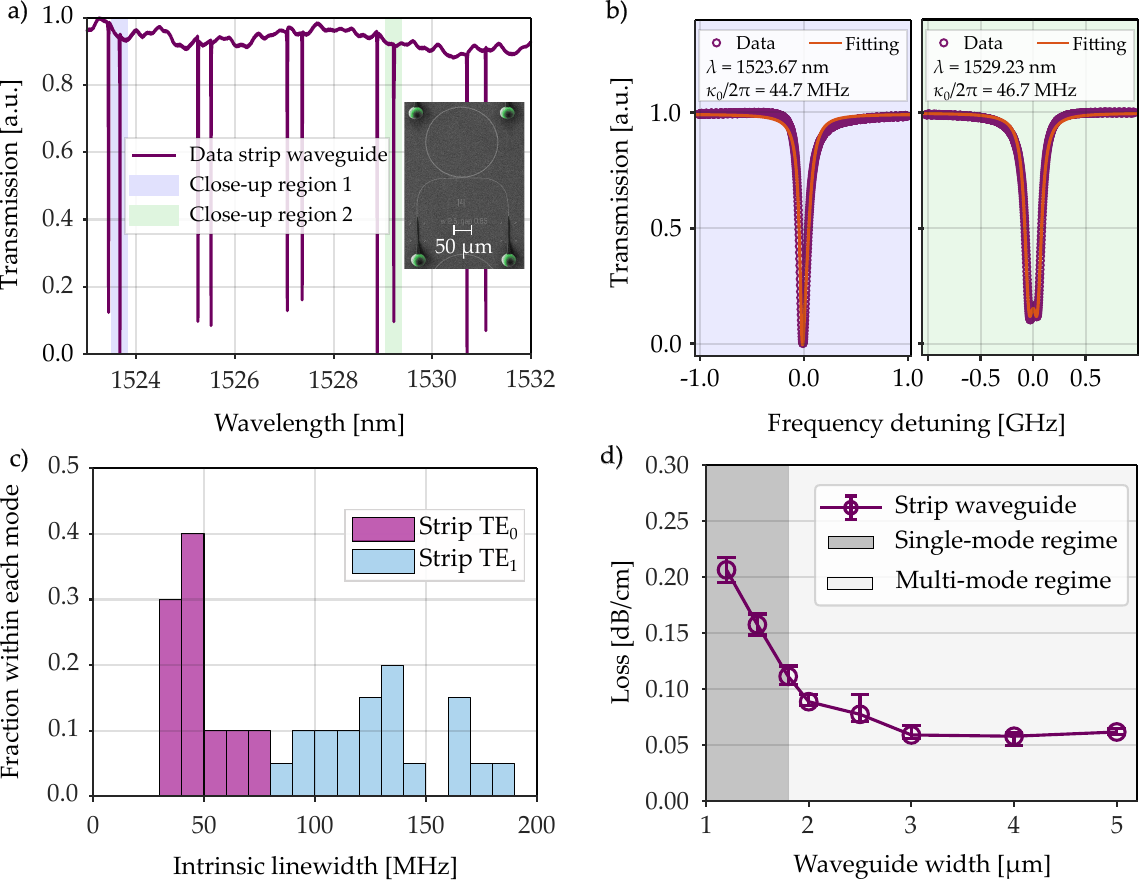}
\caption{\textit{Waveguide performance: } (a) Measured normalized spectrum with two highlighted TE$_0$ modes in a 2.5 $\mu$m-width ring resonator; the corresponding enlarged spectra are shown in (b). The inset shows a representative SEM image of the ring resonator. (b) Close-up spectra of the two TE$_0$ modes, fitted using Fano and resonance-splitting models, respectively. (c) Statistics of the measured intrinsic linewidths of the TE$_0$ and TE$_1$ modes in a 2.5 $\mu$m-width ring resonator. (d) Summary of the TE$_0$-mode propagation loss for different waveguide widths, with the single-mode and multi-mode regimes indicated. Markers show the median loss, and error bars indicate the interquartile range of the extracted resonances.}
\label{fig4}
\end{figure}

The transmission spectra obtained with unpolarized and polarized light differ in absolute value due to the polarization dependence of the 3D-printed couplers. For both designs, the transmission varies significantly with polarization, resulting in an approximate 3 dB difference between unpolarized and polarized measurements.
Consequently, when operated with a polarized optical source, the couplers are expected to exhibit an overall transmission improvement of about 3 dB across the entire measured wavelength range while maintaining their ultra-broadband response.

To characterize the dependency of propagation loss on waveguide width in the LTOI platform, ring resonators with different waveguide widths are designed and fabricated on the fully etched chip. 
Fig. \ref{fig4}(a) shows the normalized spectrum for a ring resonator with a waveguide width of 2.5 $\mu$m and a coupling gap of 850 nm. The ring radius is 100 $\mu$m. An SEM example is shown in the inset. The coupling gap between the ring and the bus waveguide is varied to determine the coupling conditions. Two resonant modes TE$_0$ and TE$_1$ are present in one free spectral range (FSR).
A close-up of two TE$_0$ resonances in Fig. \ref{fig4}(a) is shown in Fig. \ref{fig4}(b). The left and right panels show the Fano fitting and the resonance split model fitting\cite{split_fit}, respectively. The resonance splitting arises from coupling between the forward- and backward-propagating modes induced by Rayleigh back scattering, which usually happens in high-quality-factor resonators \cite{loncar_aplp}. 

Fig. \ref{fig4}(c) shows the statistics of the measured intrinsic linewidths for the TE$_0$ and TE$_1$ modes in the same ring resonator in Fig. \ref{fig4}(a). They are identified based on their distinct propagation losses, with the TE$_1$ mode showing a higher loss because of its stronger overlap with the waveguide sidewalls. The median intrinsic linewidths of the TE$_0$ and TE$_1$ modes are around 43 and 130 MHz, respectively, corresponding to propagation losses of 0.08 and 0.24 dB/cm, calculated using their respective group indices of 1.958 and 1.998. Fig. \ref{fig4}(d) summarizes the propagation loss of the TE$_0$ mode for different waveguide widths. The markers indicate the median loss extracted from the TE$_0$ resonances across the spectrum, and the error bars represent the interquartile range. The corresponding histogram statistics for each waveguide width are provided in the Supplementary Information. As the waveguide width increases, the propagation loss decreases because the optical mode volume interacting with the waveguide sidewalls decreases. Waveguide widths below 1.8 $\mu$m fall within the single-mode regime, where the median propagation loss reaches 0.11 dB/cm. For the multi-mode regime, we achieve LTOI waveguides with propagation losses below 0.1 dB/cm.

\begin{figure}[htbp]
\includegraphics[width=\linewidth]{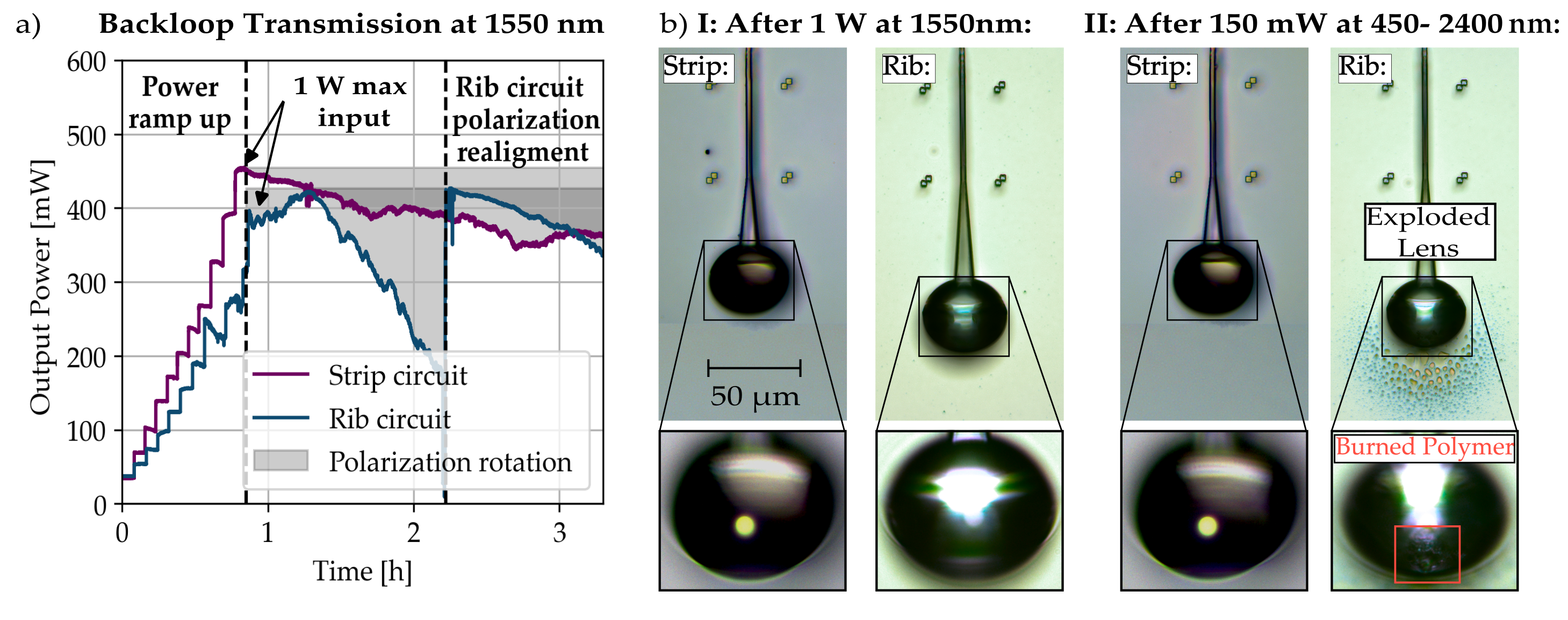}
\caption{\textit{High power damage threshold: }(a) Transmission of strip and rib backloop interfaced with TIR couplers at 1550 nm over 3 hours. In the first hour, the power is ramped-up to 1 W of input power. The polarization-rotation induced deviation from the max power is depicted in gray. The microcope images (b) are showing the coupler with a close up of the lens after the exposure with 1 W at 1550 nm (I) and 150 mW spanning from 450-2400\,nm out of the white light source (II). The couplers are capable to withstand 1W of input power at 1550nm for 2 hours without visible damage. Strip waveguide couplers can also be operated with 150 mW spanning from the visible to the infrared, whereas the rib waveguide couplers are damaged using this input power.  }
\label{fig5}
\end{figure}

To benchmark the maximum optical power with which LTOI circuitry interfaced via TIR couplers can be operated, the input laser was connected to an erbium-doped fiber amplifier (EDFA) to increase the optical power at 1550 nm. In Fig. \ref{fig5}(a), the output power of a strip and rib waveguide back-loop, interfaced with their respective TIR couplers, is plotted over a time interval of three hours. During the first hour, the power was gradually increased in 5-minute steps to evaluate whether the waveguides and couplers are capable to withstand the power up to the maximum EDFA output of 1 W. Subsequently, the system is operated at the maximum output power of the EDFA for two hours.

Apart from  polarization-induced drifts (evident from the need for polarization realignment in the rib waveguide case), the couplers exhibit stable performance under high optical power. This stability is further confirmed by microscope images of the TIR couplers shown in Fig. \ref{fig5}(b) (I), taken after exposure to the amplified laser light. For both ridge and rib waveguide couplers, no visible signs of damage are observed.

In addition, both couplers are exposed to high optical power from a white light source without the 20 dB attenuator. This  corresponds to a non-uniformly distributed power of  150 mW over the spectrum spanning from 450 nm-2400 nm. The strip waveguide coupler with IPX-Clear withstood this exposure without any observable degradation. In contrast, the transmission of the rib waveguide coupler with IPN-162 dropped to zero within one minute. The resulting damage, particularly at the lens structure, is clearly visible in the microscope image shown in Fig. \ref{fig5}(b) (II).

\section{Discussion}
The combination of broadband, low-loss couplers and waveguides enables the realization of highly efficient PICs for applications in optical communication and quantum photonics. In this work, we demonstrate the first successful interfacing of fully etched and partially etched low-loss LTOI waveguides with octave-bandwidth polymer couplers.

While previous coupling approaches based on direct laser writing have primarily focused on fully etched geometries, we extend this concept to partially etched (rib) waveguides, which are widely used in electro-optic modulators due to their enhanced light–matter interaction. To the best of our knowledge, this is the first demonstration of interfacing polymer TIR couplers with partially etched waveguides. A key challenge in such configurations is the residual slab layer, which can support parasitic modes and reduce coupling efficiency. We show that this limitation can be effectively overcome by employing a higher-index polymer material and optimizing the TIR coupler geometry. This ensures that the optical mode is transferred into the polymer waveguide rather than into the slab.
\subsection*{Coupler Performance}

Experimentally, we achieve a peak coupling loss of 0.9 dB per coupler for fully etched strip waveguides with a 1 dB bandwidth of 485 nm, and 1.25 dB per coupler for partially etched ribwaveguides with a 1 dB bandwidth of 469 nm. Both coupler types exhibit a 3 dB bandwidth exceeding an octave from 850 nm to 1750 nm making them prime candidates for SHG based squeezing applications, as both the fundamental signal and the second harmonic signal can be coupled with low-losses.

\begin{table*}[t]
\newcolumntype{Y}{>{\centering\arraybackslash}X}
\centering
\begin{threeparttable}
\caption{Performance comparison of fiber-to-chip couplers and waveguide propagation losses on LNOI and LTOI platforms. The best results for out-of-plane couplers are highlighted in bold. }
\label{tab:coupler_comparison}
\scriptsize
\renewcommand{\arraystretch}{1.25}
\setlength{\tabcolsep}{3pt}

\begin{tabularx}{\textwidth}{
>{\centering\arraybackslash}p{0.7cm}
>{\centering\arraybackslash}p{1.5cm}
Y
>{\centering\arraybackslash}p{1.5cm}
Y
Y
>{\centering\arraybackslash}p{1.7cm}
}
\toprule
\textbf{Paper}&
\makecell{\textbf{Platform/}\\\textbf{Type}} &
\makecell{\textbf{Peak coupling}\\\textbf{efficiency [dB]}} &
\makecell{\textbf{1 dB Band-}\\\textbf{width [nm]}} &
\makecell{\textbf{Fiber type}\\\textbf{(MFD)}} &
\makecell{\textbf{Coupling}\\\textbf{strategy}} &
\makecell{\textbf{Propagation}\\\textbf{loss [dB/cm]}} \\
\midrule
\multicolumn{7}{c}{\textbf{Out-of-plane couplers}} \\
\midrule

\cite{table_ref_11} & LNOI/Out-of-plane & $-1.4$  & 38  (3 dB) & Standard SMF (10.4 $\mu$m) & Chirped grating + reflector & / \tnote{a} \\

\cite{table_ref_12} & LNOI/Out-of-plane & \textbf{-0.89} & 45  & Standard SMF (10.4 $\mu$m) & Cavity-assisted grating + top metal mirror & / \\

\cite{table_ref_13} & LNOI/Out-of-plane & $-2.2$  & 81 (3 dB) & Standard SMF (10.4 $\mu$m) & Polysilicon overlay grating & / \\

\cite{table_ref_14} & LNOI/Out-of-plane & $-2.97$ & 46 & Standard SMF (10.4 $\mu$m) & Single-step apodized grating & / \\

\cite{table_ref_15} & LNOI (z-cut)/Out-of-plane & $-3.8$  & 71.7 & Standard SMF (10.4 $\mu$m) & Inverse-designed single-step grating & / \\

\cite{table_ref_16} & LTOI/Out-of-plane & $-4.6$ & $>40$  & Standard SMF (10.4 $\mu$m) & Single-step grating & 0.2 \\

\cite{table_ref_17} & LTOI/Out-of-plane & $-7.9$ & / & Standard SMF (10.4 $\mu$m) & Single-step grating & 0.09 (MM) \\

This work & LTOI/Out-of-plane & \textbf{-0.9 (fully) /-1.25 (partially)} & \textbf{485 (fully) / 469 (partially) } & Standard SMF (10.4 $\mu$m) & 2PP 3D printing & 0.11 (SM) / 0.06 (MM) \\
 \midrule
\multicolumn{7}{c}{\textbf{In-plane couplers}} \\
 \midrule
 \cite{table_ref_2} & LNOI/In-plane & $-0.75$ & $>100$ & UHNA fiber (3.2 $\mu$m) & Tri-layer converter & / \\
 \cite{table_ref_5}& LNOI/In-plane & $-0.29$ \tnote{b} & 180 (0.5 dB) & UHNA fiber (4.8 $\mu$m) & Inverse taper with silica ridge + backside removed & 0.3 considered \\

\cite{table_ref_9} & LTOI/In-plane & $-0.75$ & 400 & UHNA Fiber (6.5 $\mu$m) & Trident In-plane coupler & / \\

\cite{table_ref_10} & LTOI/In-plane & $-1.25$  \tnote{c} & / & Lensed fiber (MFD not mentioned) & Double-layer taper & 0.0595\\

\bottomrule
\end{tabularx}

\begin{tablenotes}
\footnotesize
\item[a] ``/'' indicates that the value was not reported.
\item[b] Fiber-to-air Fresnel loss of $\sim 0.15$ dB is excluded.
\item[c] Coupling efficiency is estimated by mode overlap.
\end{tablenotes}

\end{threeparttable}
\end{table*}

Table \ref{tab:coupler_comparison} compares the coupling performance and waveguide propagation losses reported for LNOI and LTOI platforms. These platforms are considered together due to their similar fabrication processes and refractive indices, while experimental results on LTOI remain comparatively scarce. The comparison focuses on state-of-the-art in-plane and out-of-plane couplers with the highest reported efficiencies.

Out-of-plane grating couplers provide significant design flexibility for on-chip architectures and enable wafer-scale characterization. However, reported coupling efficiencies range from $-7.9$ dB to $-2.97$ dB per coupler for single-step grating \cite{table_ref_14,table_ref_15,table_ref_16, table_ref_17}. This can be improved to $-0.89$ dB per coupler using advanced approaches such as metallic mirrors and reflective structures\cite{table_ref_11,table_ref_12,table_ref_13}. Nevertheless, the demonstrated bandwidth of grating couplers is generally limited to below 100 nm. In contrast, the polymer-based TIR couplers presented here achieve comparable loss performance to the best presented grating coupler \cite{table_ref_12} while exceeding the grating-coupler bandwidth by a factor of five.

The polymer-based out-of-plane couplers achieve peak coupling efficiencies comparable to state-of-the-art in-plane couplers\cite{table_ref_2,table_ref_5,table_ref_9,table_ref_10} while eliminating the need for polished chip facets and simultaneously enabling wafer-level characterization. In terms of bandwidth, they even surpass conventional in-plane couplers, exhibiting a 1 dB bandwidth exceeding 450 nm and a 3 dB bandwidth beyond 900 nm. Also, the alignment tolerances of the out-of-plane couplers is in general higher than the one of the in-plane couplers based on the mode field diameters (MFDs) used to interface the couplers.

Notably, even at wavelengths as short as 750 nm, where slab-guided modes are expected to occur, the transmission of the rib waveguide couplers decreases by only 2 dB relative to the peak value. This demonstrates the robustness of the coupling concept and indicates that the presented approach can be extended to partially etched platforms with thicker slabs or higher refractive-index contrasts, such as LNOI.

\subsection*{Low-loss and high power damage threshold LTOI circuitry}

In the present work, the LTOI waveguides are fabricated using an electron-beam resist mask for process simplicity. The resulting devices exhibit low propagation losses of 0.11 dB/cm for the single-mode waveguide and 0.06 dB/cm for the multimode waveguide. Further reduction of propagation loss is expected by employing more robust hard masks, such as the diamond-like carbon \cite{tjk_lt_nature_2024,lt_10M}, which can improve etching quality and sidewall smoothness. In addition, introducing an upper cladding can further suppress scattering loss by reducing the refractive-index contrast at the waveguide sidewalls \cite{loncar_aplp}.

In addition to broadband performance, the couplers demonstrate remarkable robustness against high optical powers. The combined system of LTOI waveguides and polymer TIR couplers sustains optical input powers of up to 1 W at 1550 nm over 2 hours operation times without any observable degradation. This highlights the robustness of both the material platform and the coupling scheme. However, a wavelength-dependent limitation is observed: in the visible regime, partially etched couplers exhibit reduced power tolerance due to increased absorption in the higher-index polymer material. This trade-off between refractive index and absorption represents an important design consideration for future implementations targeting shorter wavelengths.
\subsection*{Outlook}
Overall, the combination of ultra-broadband and low-loss coupling, compatibility with partially etched waveguides and high-power resilience positions this approach as a strong candidate for next-generation integrated photonic systems.
The high optical damage threshold is particularly advantageous for nonlinear optical processes, enabling significantly higher pump powers and thus improving the efficiency of squeezed-light generation. 

In particular, the combination of low-loss waveguides with octave-spanning coupling interfaces capable of simultaneously extracting both the fundamental and second-harmonic modes, together with fast electro-optic modulation and efficient second-harmonic-generation-assisted squeezing, provides a versatile platform for integrated quantum photonics. These capabilities could facilitate the realization of large-scale photonic quantum computing circuits by supporting low-loss state preparation, manipulation, and readout within a single integrated platform.
\section{Methods}
\subsection{Fabrication}
The waveguide is first patterned by electron-beam lithography using the AR-N 7520, and transferred to the LTOI layer by inductive-coupled plasma etching. The redeposition of the LTOI is removed by a KOH solution. Afterwards, the chip is cleaned by an RCA-1 process. In the end, the chip is annealed at 400 $^\circ$C. The RCA-1 process alone is not capable to remove the redeposition of LTOI due to Ta-rich redeposition, which typically requires higher alkalinity \cite{tjk_lt_nature_2024}. 

The polymer couplers are fabricated with a Nanoscribe Quantum X align system combined with a 63x-microscope objective in Dip-in mode. The job generation is performed with NPXPY \cite{caghanunluerCuenlueerNpxpyV0112025}, a python based open source text-only project preparation framework for Nanoprint X jobs. For the fully-etched waveguide couplers, the resin IPX-clear, for the partially-etched waveguide couplers the resin IPn-162 is used. Based on the detection of on-chip markers, the couplers are with a sub 200 nm tolerance printed a the destinated positions. After the couplers where printed, the non-polymerized liquid photoresin is washed away using Propylen Glycol Monomethyl Ether Acetat (PGMEA) and Isopropanol before blow-drying the strucutre.

\subsection{Measurement setup}
The 3D couplers are characterized using on-chip LTOI backloops. For the optical measurements (Fig. \ref{fig3}(c)), a tunable laser (Santec, TSL-770) is used to sweep the wavelength from 1480 to 1620 nm. A polarization controller is used to ensure TE-mode excitation. A multichannel fiber array with standard single-mode fibers is used to couple light into and out of the on-chip backloops.
The optical output power is measured using a photoreceiver (Newport, 2053-FS). To calibrate the reference input power the on-chip backloop structure is replaced with a single-mode fiber in order to account for the losses induced from the fiber connectors and polarization controller.
The losses from the fiber array and Fresnel reflection are included in the measured coupler transmission.

In the broadband measurements (Fig. \ref{fig3}(d)), the tunable laser is replaced by a white light source (NKT Photonics, SuperK COMPACT), and the light is collected by an OSA (Ando, AQ6317B).
In the high optical power measurements (Fig. \ref{fig5}), an EDFA (PriTel, LNHPFA-33) after the tunable laser is placed after the tunable laser to boost the optical power to 1 W right before the chip input, and a 20 dB optical attenuator is inserted after the backloop output to protect the photoreceiver. 

Similarly, the ring resonators are characterized using the same tunable laser, polarization controller, and photoreceiver. A data acquisition device is used to transfer the data from the photoreceiver signal to a computer. An attenuator is inserted after the laser output to ensure the optical power at the ring input is below 0.01 mW. 

{\small \section*{Acknowledgment} E.J. and W.P. acknowledge the funding from the Deutsche Forschungsgemeinschaft (DFG; German Research Foundation) via the Excellence Cluster “3D Matter Made to Order” (EXC-2082/1-390761711) and the Carl Zeiss Foundation through the Carl-Zeiss-Foundation-Focus@HEiKA. X.M. and W.P. acknowledge the funding from Bavarian State Government (no. Z.5-F5121.17.1/6/25, MQV project), and the Federal Ministry of Research, Technology and Space (BMFTR) (no. 13N16074, Muniqc-atom project). X.M. acknowledge funding from the European Union’s Horizon Europe research and innovation programme under the Marie Skłodowska-Curie grant agreement (no. 101205112, HEIVOM project).}


{\small \section*{Data availability}All data needed to evaluate the conclusions in the paper are present in the paper and/or the Supplementary Materials.}
\\

\par\noindent\authormark{$\dag$}These authors contributed equally to this work.\\


\bibliography{references3,sample}






\end{document}